\documentstyle[epsfig]{elsart}

\def\slashchar#1{\setbox0=\hbox{$#1$}           
   \dimen0=\wd0                                 
   \setbox1=\hbox{/} \dimen1=\wd1               
   \ifdim\dimen0>\dimen1                        
      \rlap{\hbox to \dimen0{\hfil/\hfil}}      
      #1                                        
   \else                                        
      \rlap{\hbox to \dimen1{\hfil$#1$\hfil}}   
      /                                         
   \fi}                                         %

\newcommand{\be}{\begin{equation}}  
\newcommand{\ee}{\end{equation}}
\newcommand{\ba}{\begin{eqnarray}}  
\newcommand{\ea}{\end{eqnarray}}

\newcommand{\msb}{\overline{\rm{MS}}}
\newcommand{\bea}{\begin{eqnarray}}
\newcommand{\eea}{\end{eqnarray}}

\newcommand{\dsl}{\stackrel{\leftarrow}{\slashchar D}}
\newcommand{\dsr}{\stackrel{\rightarrow}{\slashchar D}}

\newcommand{\cals}{{\cal S}}
\newcommand{\calo}{{\cal O}}

\newcommand{\pslash}{p\!\!\!/\,}

\newcommand{\as}{\alpha_s}

\def\ct#1{{\cal #1}}
\def\dfrac#1#2{{\displaystyle {#1 \over #2}}}

\newcommand{\Pj}{\mbox{I}\!\!\mbox{P}}

\newcommand{\Tr}{\mbox{Tr}\;}

\newcommand{\RI}{\mbox{\scriptsize RI}}

\newcommand{\msbar}{\overline{\mbox{\scriptsize MS}}}
\newcommand{\MSbar}{\overline{\mbox{MS}}}

\begin{document}

\rightline{Edinburgh 98/10}
\rightline{Rome1-1199/98}
\rightline{ROM2F/98/24}
\rightline{SNS/PH/1998-016}

\begin{frontmatter}

\title{The QCD Chiral Condensate from the Lattice}

\author{L.~Giusti}
\address{Scuola Normale Superiore, P.zza dei Cavalieri 7, I-56100 Pisa, 
Italy\\  INFN-Sezione di Pisa, I-56100 Pisa, Italy}
\author{F.~Rapuano}
\address{INFN-Sezione di Roma, c/o Dipartimento di Fisica,\\
Universit\`a di Roma \lq La Sapienza\rq, P.le A. Moro 2,\\
I-00185 Roma, Italy.}
\author{M.~Talevi}
\address{Department of Physics \&\ Astronomy, University of Edinburgh\\
         The King's Buildings, Edinburgh EH9 3JZ, UK}
\author{A.~Vladikas}
\address{INFN-Sezione di Roma II, c/o Dipartimento di Fisica,\\
Universit\`a di Roma \lq Tor Vergata\rq, Via della Ricerca Scientifica 1,\\
I-00133 Roma, Italy.}

\begin{abstract} 
We determine the chiral condensate from simulations of quenched lattice QCD
with Wilson fermions. Our measurements have been obtained with high statistics
at three values of the gauge coupling, corresponding to UV cutoffs in the range
$2 - 4$~GeV. Several improvements have been made with respect to earlier lattice
computations. The most important are the non-perturbative renormalization of
the condensate, the use of the tree-level improved Clover action and the
reduction of the systematic error due to uncertainties in the lattice
calibration. Our result for the chiral condensate in the $\msbar$ scheme is
\\ $\langle \bar \psi \psi \rangle^{\msbar}(\mu =~2~{\rm GeV})~=~ 
-~0.0147(8)(16)(12)~{\rm GeV}^3 =~-~[245(4)(9)(7)~{\rm MeV}]^3$ \\
where the first error is statistical, the second is due to the 
non-perturbative renormalization and the third due to the lattice
calibration.
\end{abstract}

\end{frontmatter}

\vfill
\centerline{PACS: 11.15.H, 11.30.Rd, 12.38.Gc 12.39.Fe and 14.40.Aq}

\newpage
\clearpage 

\section{Introduction}

QCD has a remarkably small mass gap. Its pattern is described by a nearly
degenerate isospin multiplet containing pseudoscalar mesons, but not their
partners of opposite parity (vector mesons). This implies that
chiral symmetry is spontaneously broken in massless QCD. This should be
signalled by the non-vanishing of some order parameter (for a given
order parameter this is a sufficient but not necessary condition for
symmetry breaking). The chiral condensate is such an order parameter.
The standard expectation is that it does not vanish in the chiral limit.
However, it cannot be {\it a priori} excluded (see ref.~\cite{gl} and, more
recently, \cite{js}) that it might tend to zero in this limit, while chiral 
symmetry breaking is nevertheless realized through the non-vanishing of the
pion decay constant. A more general option is that the chiral condensate has a
small non-zero value \cite{js}.

The intrinsically
non-perturbative nature of the chiral condensate implies that its determination
from first principles is not straightforward. QCD sum rules and lattice
computations have been used in order to estimate its value. 
The lattice regularization is particularly well suited for the computation of
low-energy non-perturbative quantities from
first principles. Although lattice measurements are
affected by a series of approximations (finite configuration ensemble, finite
volume and lattice spacing, non-zero quark mass, quenched approximation),
they can be controlled and systematically improved, at least in principle.
Two of these sources of error have the most negative impact on the credibility
of the lattice determination of the chiral condensate, namely the quenched
approximation and the chiral extrapolation of data computed at non-zero
quark masses. Quenching is destined to stay with lattice computations for some
time, since realistic simulations with light sea quark masses are still
unaccessible to present-day computers. Quenched simulations with fairly light
valence quarks are more accessible, but the zero quark limit is elusive
for hadronic quantities\footnote{
Note however, that it is possible to measure the chiral condensate
with staggered fermions at zero quark mass, using the Banks-Casher formula
\cite{bc} (see for example refs.~\cite{gla}). More recently, the
Shr\"odinger Functional formalism has been used for Wilson fermions in order
to obtain lattice renormalization constants at zero quark mass \cite{lusch1}.}.

Interestingly, the two approximations (quenching and chiral extrapolations) are
interwoven by the presence of quenched chiral logarithms in many physical
quantities, such as the chiral condensate and the pion mass (but not
in the pion decay constant, in the isospin limit) \cite{qchlogs}. There has
been some debate recently on the evidence for quenched chiral logarithms in
lattice results obtained with light quark masses \cite{kimsin}.
However, for the bulk of quenched lattice results, obtained at
the strange quark mass region, the power
dependence on the quark mass dominates the logarithms. In such simulations
many physical quantities known from experiment, including those related to the
chiral condensate (e.g meson masses, decay constants), have been determined
with a 5-15\% accuracy (see for example ref.~\cite{mio1}). Also, the recent
lattice determinations of the quark masses \cite{m_q_APE} agree with the sum
rules predictions. The good agreement of many quenched results with experiment
\footnote{There are several exceptions, such as the disagreement 
of the experimental and the lattice result (in the quenched approximation) of 
the decay constant ratio $f_K/f_\pi$.},
obtained in a quark mass region where chiral singularities can be ignored,
implies that the regular dependence of these observables on the quark mass
is adequately reproduced in the quenched approximation.

Given the above observations, we have performed an extensive quenched lattice
computation of the chiral condensate, with Wilson fermions, using the data
accumulated by the APE collaboration in the last few years. The valence quark
masses used in these simulations are in the range of the strange flavour.
Several important improvements have been possible compared to earlier 
computations:
\begin{itemize}
\item{We have improved our statistics by computing the condensate on ensembles
of $\ct{O}(100)$ gauge field configurations. Most previous computations with Wilson
fermions were performed on $\ct{O}(10)$ configurations.}
\item{Reliable measurements have been performed at two values of the lattice
gauge coupling $g_0^2$, corresponding to UV cutoffs of $\sim 2$~GeV and 
$\sim 3$~GeV, using both the Wilson and the tree-level
SW-Clover\cite{sw,heatlie} actions. Results obtained with the Wilson action
at fixed UV cutoff (i.e. fixed lattice spacing $a$) suffer from $\ct{O}(a)$
discretization errors; those obtained in the tree-level Clover formalism
suffer from $\ct{O}(a g_0^2)$ errors. Thus we can study the influence of
finite lattice spacing effects. We also present results at a third value of the
gauge coupling (corresponding to an UV cutoff of $\sim 4$~GeV), which are
however
somewhat unreliable, due to the small physical size of the lattice.}
\item{Different determinations of the chiral condensate (expressed in lattice
units) have been implemented. They all derive from the same Ward
Identity(WI),
obtained from the lattice action, and are equivalent only in the chiral and
continuum limits. They are implemented, however, at non-zero lattice spacing and
quark masses and subsequently extrapolated to the chiral limit. Thus, by
comparing the chiral condensate from different determinations, we can
cross-check the reliability of the extrapolations.}
\item{The systematic errors arising from the chiral extrapolation and the
uncertainties in the determination of the lattice spacing are amplified
in the standard computation of the chiral condensate. The chiral extrapolation
amplifies the error because we extrapolate from data obtained in the
strange quark region. The uncertainty in the determination of the lattice
spacing is amplified because the condensate is a dimension-3 matrix element.
We have obtained the lattice condensate
in physical units in a way that avoids these problems.}
\item{The chiral condensate measured on the lattice  at a given fixed lattice
spacing is a bare quantity which must be renormalized. In previous studies,
its renormalization constant was calculated in lattice Perturbation Theory (PT)
at 1-loop. We will show that the perturbatively renormalized chiral condensate
suffers from large $\ct{O}(g_0^4)$ corrections, which render its determination
unreliable (an identical conclusion was drawn for the quark mass in
ref.~\cite{m_q_APE}).
In the present work, we have used non-perturbative (NP) estimates
of the required renormalization constant, which are known in the Regularization
Independent (RI) renormalization scheme (also known as the MOM
scheme). This removes lattice PT completely from the renormalization
procedure. We only use continuum PT in order to transform the results from the
RI to the intrinsically perturbative $\MSbar$ scheme. Also,
RG running has been done in PT in order to express all results in the
conventional scale of 2~GeV. All PT coefficients used in this work are at 
NNLO.}
\end{itemize}

Our best estimate for the chiral condensate agrees with the earlier lattice
determinations of refs.~\cite{mm,daniel} with greatly improved statistical
accuracy. For the
first time we are also able to estimate the systematic uncertainties in a
reliable way. Our result is also in excellent agreement with the one obtained
from sum rules  (see \cite{dn} and references therein).

We now outline the contents of this paper. In sect.~\ref{sec:defs} we fix our
notation and give the basic definitions. In sect.~\ref{sec:renormcc} we discuss
issues related to our renormalization strategy. In particular we review the
most important properties of the NP renormalization of the lattice condensate
(see Appendix~\ref{app:npm} for details). We also review the connection between
the RI and $\MSbar$ renormalization schemes (details can be found in
Appendix~\ref{app:match}). In sect.~\ref{sec:wwi} we discuss how the chiral
condensate can be determined from the lattice regularization with Wilson
fermions.
Chiral symmetry, explicitly broken by the Wilson fermionic action, is recovered
in the continuum and chiral limits through the implementation of lattice WIs;
see ref.~\cite{ksboc}. With the aid of such a WI we obtain the proper lattice
definition of the chiral condensate which, for Wilson fermions, is
subject to a power subtraction. Moreover, from this WI we derive the well known
Gell-Mann--Oakes--Renner (GMOR) relation \cite{gmor} for the condensate. Some
technicalities related to the WIs with Clover fermions are discussed in
Appendix~\ref{app:cwi}. In sect.~\ref{sec:meascc} we present our results and
compare them to those of previous computations of the chiral condensate with
Wilson fermions\footnote{
No comparisons were made with results obtained with Kogut-Susskind
fermions, which are usually presented unrenormalized.
We also point out that this work is limited to zero-temperature QCD.
For a recent review on lattice studies of chiral symmetry restoration at
finite-temperature QCD see ref.~\cite{laer}.}.
Our conclusions are summarized in sect.~\ref{sec:concl}.

\section{Definitions and generalities}
\label{sec:defs}

In this section we define the quantities of interest to this paper, in order
to fix our notation. We work in the lattice regularization scheme proposed by
Wilson \cite{wils1}. The partition function of the theory is
\be
Z = \int \ct{D} \psi \ct{D} \bar \psi \ct{D} U_\mu
\exp \left( -S_g - S_f \right)\; ,
\label{eq:pf}
\ee
where the gluonic action is given by
\be
S_g = \dfrac{6}{g_0^2}\sum_{P}
\left\{1-\dfrac{1}{6} Tr \bigl[ U_P + U_P^{\dagger} \bigr]\right\}
\label{eq:sg}
\ee
and the fermionic action by 
\bea
S_f & = & -a^4 \sum_{x,\mu} \dfrac{1}{2a} 
 \bigl[ \bar \psi (x) (1 - \gamma_\mu) U_\mu (x) \psi (x+\mu)
 + \bar \psi (x+\mu) (1 + \gamma_\mu) U^\dagger_\mu (x) \psi (x) \bigr]
\nonumber \\
&& + a^4 \sum_x \bar \psi (x) (M_0 + \dfrac{4}{a}) \psi (x)\; . 
\label{eq:sf}
\eea 
In standard notation, $a$ is the lattice spacing, $\psi (x)$ is the quark
field (flavour indices are implied), $U_P$ is the Wilson
plaquette, $U_\mu(x)$ is the lattice gauge link and $g_0$ is the bare
coupling constant. The diagonal bare mass matrix is denoted by $M_0$.
In most of this work we will assume mass degeneracy; i.e. all elements of
$M_0$ are equal to $m_0$. As is well known, for Wilson fermions, the 
quark mass, besides a multiplicative renormalization, is also subject to an 
additive one. Defining $m = m_0 - m_C$, the renormalized is
\be
m_R = Z_m m = Z_m [ m_0 - m_C ]
\label{eq:mrzm}
\ee
with $Z_m$ the multiplicative renormalization constant. The chiral limit is
then $m\rightarrow 0$; i.e. $m_0 \rightarrow m_C$. At tree-level, $m_C = -4/a$.

We define the scalar and pseudoscalar densities as:
\bea
S^f(x) &=& \bar \psi(x) \dfrac{\lambda^f}{2} \psi(x)
\nonumber \\
P^f(x) &=& \bar \psi(x) \gamma_5 \dfrac{\lambda^f}{2} \psi(x)
\label{eq:sp}
\eea
and the vector and axial currents as
\bea
V_\mu^f(x) &=& \bar \psi(x) \gamma_\mu \dfrac{\lambda^f}{2} \psi(x)
\nonumber \\
A_\mu^f(x) &=& \bar \psi(x) \gamma_\mu \gamma_5 \dfrac{\lambda^f}{2} \psi(x)\;
,
\label{eq:va}
\eea
where $\lambda^f/2$ are the generators of the $SU(N_f)$ flavour group in the 
fundamental representation (e.g. Pauli matrices for $N_f=2$, Gell-Mann
matrices for $N_f=3$ with $f = 1,\dots,N_f^2-1$). These generators satisfy: 
\bea
&& Tr[\lambda^f \lambda^g ] = 2 \delta^{fg} \nonumber \\
&& \left[ \dfrac{\lambda^f}{2}, \dfrac{\lambda^g}{2}\right]
=i f^{fgh}\dfrac{\lambda^h}{2} \nonumber \\
&& \left\{\dfrac{\lambda^f}{2}, \dfrac{\lambda^g}{2}\right\}
=d^{fgh}\dfrac{\lambda^h}{2}+\dfrac{\delta^{fg}}{N_f}\,\lambda^0\; ,
\label{eq:ds}
\eea
where $f^{fgh}$ are the $SU(N_f)$ structure constants, $d^{fgh}$ are symmetric
coefficients and $\lambda^0$ represents the $N_f \times N_f$ identity matrix. 
We will also consider the corresponding singlet bilinear scalar operator
\be
S^0(x) = \bar \psi(x) \lambda^0 \psi(x) = \bar \psi(x) \psi(x)\; .
\label{eq:spva0}
\ee
The operators defined above are to be understood as bare lattice 
quantities\footnote{
This means that all integrals $\int d^4x$ which appear below, are
to be interpreted as sums ($a^4 \sum_x$) which run over all lattice sites,
labelled by $x$; $\delta(x)$ stands for $a^{-4} \delta_{x0}$, etc. The use
of integrals instead of sums and Dirac functions instead of Kronecker symbols
is a question of notational convenience and should be transparent.}.
They are subject to multiplicative
renormalization. The renormalization of a generic bilinear operator
$O_\Gamma^f = \bar \psi \Gamma (\lambda^f/2) \psi$ is given by
($\Gamma$ stands for any Dirac matrix)
\be
\hat O_\Gamma(g_R^2,m_R,\mu) = \lim_{a \rightarrow 0}
[Z_O(g_0^2,a\mu) O_\Gamma(g_0^2,m,a)]\; ,
\label{eq:zodef}
\ee
where $g_R$ is the renormalized gauge coupling and $\mu$ the renormalization 
scale. Note that in the bare operator the bare mass $m_0$ has been traded off 
for the more convenient (from the point of view of the chiral limit)
subtracted  mass $m$. Renormalized quantities will be denoted in general by
``hats"; e.g. the renormalized quark propagator is
\be
\hat \cals (p) = \lim_{a \rightarrow 0}[Z_q \cals (p)]\; ,
\label{eq:cp}
\ee
where $Z_q^{1/2}$ is the quark field renormalization. 

The properties of the renormalization constants of the operators of
eqs.~(\ref{eq:sp}) and (\ref{eq:va}), obtained in refs.~\cite{ksboc}
(for a recent treatment see \cite{clv}), are as follows:
The renormalization constants $Z_S$
and $Z_P$ of the scalar and pseudoscalar densities are logarithmically
divergent; thus they depend on the coupling $g_0^2$ and the renormalization
scale in lattice units $a\mu$. Those of the vector and axial currents ($Z_V$
and $Z_A$ respectively) are finite; thus they depend only on the coupling
$g_0^2$. A mass independent renormalization scheme is implied.

\section{Renormalization scheme dependence of the chiral condensate}
\label{sec:renormcc}

In this section we discuss some issues related to the choice of renormalization
scheme for the chiral condensate. The renormalization procedure which gives a
finite condensate starting from its bare (lattice) counterpart is postponed
until the next section. Here we will concentrate on how to connect the
condensate in the RI renormalization scheme, obtained
non-perturbatively, to the one in the $\MSbar$ scheme. This is necessary 
in view of
the fact that the RI scheme is most convenient for lattice computations,
as discussed below, whereas the $\MSbar$ scheme is conventionally
preferred when expressing ``final'' physical results.

Let us sketch step by step the renormalization procedure for the chiral
condensate.
The breaking of chiral symmetry by the Wilson term of the fermionic action
$S_f$ induces some subtleties in the renormalization of the lattice
condensate. These have been dealt with in ref.~\cite{ksboc} and will be fully
exposed in sec.~\ref{sec:wwi}. Here we simply anticipate the basic result:
by using lattice WIs and by requiring that in the continuum and chiral
limit the nominal WIs are recovered, we obtain that the correct
renormalization of the lattice chiral condensate is given by
\be
\langle \bar \psi \psi \rangle (\mu) = Z_P(\mu a) \left[
\langle \bar \psi \psi \rangle (a) - b_0 \right]\; ,
\label{eq:chcondRI}
\ee
where $\langle \bar \psi \psi \rangle (a) = \langle S^0 \rangle$ stands
for the bare chiral condensate, computed non-perturbatively on the lattice, 
and $b_0$ is a cubicly diverging subtraction.
Note that $Z_P$ is the renormalization of the non-singlet pseudoscalar density.
Its determination from 1-loop lattice perturbation theory (see
refs.~\cite{z2pert,gabri,ari} for the perturbative renormalization of quark
bilinears) suffers from large uncertainties due to the presence of
``tadpole'' diagrams \cite{lep_mac}. In order to avoid this problem
non-perturbative  renormalization techniques have been developed
\cite{lusch1,mm,gri,NPM,clvwi} consisting in computing the renormalization
constants at fixed UV cutoff (ranging, in present-day simulations, in the 
region $a^{-1} \simeq 2 - 4$~GeV). 
This is a tradeoff between higher order corrections (present in PT) and
lattice artifacts (present in all non-perturbative methods). The latter,
however, can be systematically controlled by improving the action and operators
in the spirit of Symanzik's work \cite{lusch1,sw,heatlie,sym}. Thus in
principle non-perturbative evaluations of the renormalization constants are 
preferable. The finite renormalization constants ($Z_V$, $Z_A$ and the
ratio $Z_P/Z_S$) can be obtained non-perturbatively with the help of WIs
(see \cite{lusch1,mm,ksboc,clv,gri,clvwi}). All renormalization
constants (including the diverging ones $Z_S$ and $Z_P$) can be obtained
non-perturbatively with the help of the more recent ``Non-Perturbative'' (NP)
method of ref.~\cite{NPM}. Results of the NP renormalization constants of quark
bilinear operators can be found in refs.~\cite{NPM,schier,Z_bl_tli}.

In practice, the necessity of non-perturbative renormalization
was first demonstrated in the measurement of several B-parameters
of the $\Delta S = 2$ four-fermion operators; see refs.~\cite{znp4f}.
More recently, it has also been shown in ref.~\cite{m_q_APE} that
the discrepancy, observed in the past between quark masses extracted from the
Vector and Axial WIs, was due to the very poor determination of $Z_P$ in
PT. Use of the NP value led to an excellent agreement between the two
determinations. The same conclusions are drawn in the present work for the
chiral condensate.

We will therefore use the NP value of $Z_P$ in the present work. The underlying
theory and method of computation has been fully exposed in \cite{NPM}; it
consists
in applying the RI (or MOM) scheme on the lattice. The amputated Green function
of the insertion of the operator $P$ in the quark propagator
is computed at fixed external momentum in the deep Euclidean region. The
Green function is subsequently projected by a ``suitable projector'' and
the RI renormalization condition fixes the renormalized projected Green 
function be equal to its tree-level value at a given
scale $\mu$. This condition can then be solved for $Z_P$; the result
is a fully non-perturbative estimate of the renormalization constant at fixed
UV cutoff. For this procedure to be reliable $\mu$ must satisfy the conditions
$\mu \ll \ct{O}(a)$ (in order to avoid discretization errors) and 
$\mu\gg \Lambda_{QCD}$. The former bound is approximately satisfied in
practical simulations; we will be using the renormalization constants
of ref.~\cite{Z_bl_tli} at the scale $\mu \simeq a^{-1}$.
The latter bound is necessary for two reasons.
(i) The RI scheme imposes that the renormalized operator transform as an
irreducible representation of the chiral group. This is true only at large
scales, where chiral symmetry breaking effects are negligible. (ii) At large
scales we avoid higher order corrections in the continuum perturbative
expansion in which the RI-$\MSbar$ matching coefficient is calculated;
see below. Unlike $\MSbar$, the RI renormalization
condition does not depend on the regularization chosen, hence its name,
Regularization Independent (RI).
The explicit formulae of the RI renormalization condition and a
discussion of some subtleties related to wavefunction renormalization can
be found in Appendix~\ref{app:npm}.

Thus the chiral condensate of eq.~(\ref{eq:chcondRI}), renormalized with the NP
method outlined above at scale $\mu \simeq a^{-1}$,
is expressed in the RI scheme. In order to enable
comparisons with other methods, we must convert our result to the $\MSbar$
scheme\footnote{
The continuum perturbative calculations related to the $\MSbar$
results quoted throughout this work have been performed in the NDR
regularization scheme.}.
This is expressed in terms of a perturbative finite matching coefficient
\be
\langle \bar \psi \psi \rangle ^{\msbar} (\mu) = \Delta Z^{\msbar,\RI}
\langle \bar \psi \psi \rangle^{RI} (\mu)\; .
\label{eq:chcondm}
\ee
The matching coefficient $\Delta Z^{\msbar,\RI}$, known in PT to NNLO
(see refs.~\cite{Z_bl_tli} and~\cite{fravit}), is given in
Appendix~\ref{app:match}. Note that
the renormalization constants obtained in the RI scheme are in
general gauge dependent (we work in the Landau gauge). This dependence
is cancelled by the gauge dependence of $\Delta Z^{\msbar,\RI}$.

In order to express our result at the conventional
scale of $\mu^\prime = 2$~GeV, we use perturbative RG running
\be
\langle \bar \psi \psi \rangle^{\msbar} (\mu^\prime)
= \dfrac{c^{\msbar}_S(\mu^\prime)}{c^{\msbar}_S (\mu)}
\langle \bar \psi \psi \rangle^{\msbar} (\mu)\; .
\label{eq:rgrun}
\ee
Finally, it is convenient to express the chiral condensate in a scale
independent way. We define the RGI chiral condensate as
\be
\langle \bar \psi \psi \rangle^{\rm RGI} = \dfrac{1}{c^{\msbar}_S (\mu)}
\langle \bar \psi \psi \rangle^{\msbar} (\mu)\; .
\label{eq:rginv}
\ee
This quantity does not depend neither on the scale $\mu$ nor on the
renormalization scheme. The evolution coefficient $c^{\MSbar}_S (\mu)$ is
known to NNLO; see Appendix~\ref{app:match}.

\section{The chiral condensate from lattice WIs with Wilson fermions}
\label{sec:wwi}

In this section we obtain the proper definition
of the chiral condensate with Wilson fermions from lattice WIs. The subject of chiral
symmetry breaking by the Wilson term and its restoration in the continuum
limit with the aid of WIs is well understood (see the original works
\cite{ksboc}, a recent review \cite{clv} and a recent theoretical treatment
of the subject \cite{testag}). Thus we only present in some detail 
those results essential to the understanding of the lattice
determination of the chiral condensate. First we shall briefly digress to
vector WIs in order to demonstrate that the quark mass renormalization is
the inverse of the renormalization of the scalar density. Then we will
obtain the proper definition of the lattice chiral condensate and its
renormalization using axial WIs from which the Gell-Mann--Oakes--Renner
relation \cite{gmor} will be derived on the lattice.

The local $SU_L(N_f)\times SU_R(N_f)$ chiral transformations of the fermionic
fields are defined as follows:
\bea
&& \delta \psi(x)  =  i \left[ \alpha_V^f (x) \dfrac{\lambda^f}{2} +
\alpha_A^f (x) \dfrac{\lambda^f}{2} \gamma_5 \right] \psi (x)
\nonumber \\
&& \delta \bar \psi (x) = -i \bar \psi (x) \left[ \alpha_V^f (x)
\dfrac{\lambda^f}{2} - \alpha_A^f (x) \dfrac{\lambda^f}{2} \gamma_5 \right]\; .
\label{eq:vtr}
\eea
With degenerate bare quark masses $m_0$, the global vector transformations are
a symmetry of the action.
From the corresponding local transformations (eqs.~(\ref{eq:vtr}) with
$\alpha_A^f = 0$) vector WIs can be derived, which indicate that the conserved
lattice vector current is a point-split operator. The no-renormalization
theorem applies to this operator. On the other hand, the lattice vector local
current of eq.~(\ref{eq:va}) is subject of a finite renormalization
$Z_V (g_0^2) \ne 1$, due to its non-conservation on the lattice.
A detailed treatment of these results can be found in
refs.~\cite{ksboc,clv}. Here we illustrate the point by writing down a typical
lattice vector WI. For simplicity, the flavour $f$ of the vector variation of
eq.~(\ref{eq:vtr}) is chosen to be non-singlet and non-diagonal (e.g. 
the lowering operator $\lambda^f= \lambda^1 - i \lambda^2$).
The two quark masses related to this flavour are denoted by $m_1$ and $m_2$
(i.e. they are non-degenerate). Then we obtain the WI
\bea
\dfrac{\delta}{\delta \alpha_V^f(x)} \langle V^g_\nu(0) \rangle = 0
\Longleftrightarrow \nonumber \\
\label{eq:vwi}
Z_V \sum_\mu \nabla_x^\mu \langle V^f_\mu(x) V^g_\nu(0) \rangle
= (m_2 - m_1) \langle S^f(x) V^g_\nu(0) \rangle\; ,
\eea
where $a \nabla_x^\mu f(x) = f(x) - f(x-\mu)$ is an asymmetric lattice
derivative. We have imposed $x \ne 0$, in order to avoid contact terms.
We can now multiply both sides of the WI by $Z_V$ and require that the
renormalized quantities obey the nominal vector WI. This implies that
the product of the quark mass and the scalar density is renormalization
group invariant. In other words, the renormalization of the scalar density
$S(x)$ is the inverse of the multiplicative renormalization of the quark mass:
\be
Z_S = Z_m^{-1}\; .
\label{eq:zsm}
\ee
\par
Far less straightforward is the implementation of the axial symmetry with
Wilson fermions, because of the presence of the chiral symmetry breaking
Wilson term in the action. The basic idea is that, by imposing suitable
renormalization conditions, PCAC is recovered in the continuum. Consider
the axial WI, obtained from eq.~(\ref{eq:vtr}) with $\alpha_V^f = 0$, arising
from the variation of the non-singlet pseudoscalar density. For $f,g \ne 0$ we
have:
\bea
\dfrac{\delta}{\delta \alpha_A^f(x)} \langle P^g(0) \rangle = 0
\Longleftrightarrow \nonumber \\
\label{eq:awi}
- \delta(x) \delta^{fg} \dfrac{1}{N_f} \langle S^0(0) \rangle =
\sum_\mu \nabla_x^\mu \langle A^f_\mu(x) P^g(0) \rangle
- 2 m_0 \langle P^f(x) P^g(0) \rangle \\
- \langle X^f(x) P^g(0) \rangle \; .\nonumber
\eea
The l.h.s. of this equation is the variation of the operator which vanishes
when $x \ne 0$. The r.h.s. is the variation of the action from which
the lattice version of the standard PCAC relation is recovered.
The operator $X^f$ in the above WI arises from the variation of the Wilson
term under the axial transformation\footnote{
The original WI, obtained as described above, involves a
point-split axial current. This current,
expanded in powers of the lattice spacing $a$, is equal to the local
current of the above WI plus terms which are absorbed in a redefinition of
$X^f$ (see refs.~\cite{ksboc,clv} for details).}.
Unlike the vector current case, $X^f$ cannot be cast in the
form of a four-divergence. It is a dimension-4 operator which, in the naive
continuum limit vanishes, being of the form ($a \times$dimension-5 operator).
However, it has divergent matrix elements
beyond tree-level. Its mixing with operators 
of equal and lower dimensions can be expressed 
as follows~\cite{ksboc}:
\be
\overline X^f(x) = X^f(x) + 2 {\overline m } P^f(x)
+ (Z_A -1) \nabla_x^\mu A^f_\mu(x)\; ,
\label{eq:xbar}
\ee
where naive dimensional arguments tell us that the mixing constant
$Z_A(g_0^2)$ is finite, whereas $\overline m(g_0^2,m_0)$ 
diverges linearly like $a^{-1}$. Logarithmic divergences have been shown
to be absent at all orders in PT \cite{curci}. The two mixing constants 
$Z_A$ and $\overline m$, and consequently $\overline X^f(x)$, are defined 
by requiring that the renormalized axial current satisfies the nominal 
(continuum) axial WI. This requirement is satisfied upon demanding that,
in the continuum limit and for vanishing quark mass, the correlation functions
of $\overline X^g$ vanishes except for localized contact terms. For the
insertion of $\overline{X}^f$ in correlation functions such as
$\langle \overline{X}^f(x) P^g(0) \rangle$ the contact terms, by flavour
symmetry, must have the form
\bea
\langle \overline{X}^f(x) P^g(0) \rangle = \dfrac{1}{N_f} b_0 \delta(x) 
\delta^{fg} + b_1 \delta^{fg} \Box \delta (x)\; ,
\label{eq:xct}
\eea
where the last term on the r.h.s. is a Schwinger term (it vanishes upon
integration over $\int d^4 x$). Combining eqs.~(\ref{eq:awi}), (\ref{eq:xbar})
and (\ref{eq:xct}) we arrive at
\bea
\label{eq:wicc}
Z_A \nabla^\mu_x \langle A^f_\mu(x) P^g(0) \rangle
- 2 (m_0 - \overline m) \langle P^f(x) P^g(0) \rangle 
- b_1 \delta^{fg} \Box \delta (x)
\\
= -\delta(x) \delta^{fg} \dfrac{1}{N_f} \left[ \langle S^0 \rangle 
- b_0 \right]\; .
\nonumber
\eea
By multiplying both sides of the above by $Z_P$, we obtain a completely finite
expression, which is the renormalized WI in the continuum.
The $b_1$-subtraction on the l.h.s. compensates the contact terms arising in
the correlation functions $\nabla^\mu_x \langle \hat A_\mu^f(x) \hat P^g (0)
\rangle$ and $\langle \hat P^f(x) \hat P^g(0) \rangle$, when $x$ is close to
the origin. Otherwise, we have recovered the nominal WI, which is now satisfied
by renormalized operators. This implies the following useful results:\\
(1) The product $\hat A^f_\mu = Z_A A^f_\mu$ must be interpreted as the
renormalized axial current ($Z_A(g_0^2) \ne 1$ is a finite renormalization).\\
(2) The chiral limit is to be defined as the value $m_C$ of $m_0$
for which the difference $m_C - \overline m (g_0^2,m_C)$ vanishes. We see
from eq.~(\ref{eq:wicc}) that the product $(m_0 -\overline m) P^f$ is
renormalization group invariant. Thus, we can write the renormalized quark
mass as:
\be
m_R = \overline Z_m [ m_0 - \overline m(m_0)]\; ,
\label{eq:mr2}
\ee
and obtain for the renormalization of the pseudoscalar density
\be
Z_P = {\overline Z}_m^{-1} \; .
\label{eq:zpm}
\ee
Combining eqs.~(\ref{eq:mrzm}), (\ref{eq:zsm}), (\ref{eq:mr2}) and
(\ref{eq:zpm}), we obtain for the ratio $Z_P/Z_S$
\be
\dfrac{Z_P}{Z_S} = \dfrac{m_0 - \overline m(m_0)}{m_0-m_C}\; .
\label{eq:zszpm}
\ee
(3) Two useful expressions are derived from eq.~(\ref{eq:wicc}), when 
$x \ne 0$ (i.e. all terms proportional to $\delta(x)$ vanish).
The first expression consists simply in the definition of the ratio
(for $f=g$):
\be
\label{eq:2rho}
2 \rho = \dfrac{2 (m_0 - \overline m)}{Z_A} =
\dfrac{\int d^3 x \nabla^0_x \langle A_0^f(x) P^f(0) \rangle}
      {\int d^3 x \langle P^f(x) P^f(0) \rangle}\; ,
\ee
where the spatial derivatives of the divergence of the axial current
vanish under the integration.
This quantity, once $Z_A$ is known, can be used for the determination of
$(m_0 - \overline m)$. The second useful expression arises upon applying
the LSZ reduction formula:
\be
Z_A \nabla^\mu_0 \langle 0 \vert A^f_\mu(0) \vert P \rangle
= 2 (m_0 - \overline m) \langle 0 \vert P^f(0) \vert P \rangle\; ,
\label{eq:mewi}
\ee
where $\vert P \rangle$ is the lowest lying pseudoscalar state (in the chiral
limit this is the Goldostone boson).
Both equations are lattice versions of the standard PCAC relation.\\
(4) Last but not least, the $b_0$-subtraction on the r.h.s. of
eq.~(\ref{eq:wicc}) implies that the proper definition of the lattice chiral
condensate with Wilson fermions, determined by axial WIs, is given by
\be
\langle \bar \psi \psi \rangle_{sub} =
\left[ \langle S^0 \rangle - b_0 \right]\; .
\label{eq:ccs}
\ee
The above ``subtracted chiral condensate" is a logarithmically divergent
quantity, which is multiplicatively renormalized by $Z_P$:
\be
\langle \bar \psi \psi \rangle = 
Z_P \left[ \langle S^0 \rangle - b_0 \right]\; .
\label{eq:cc}
\ee
\par
To obtain an expression from which the chiral condensate can be
computed in a convenient way, we need to integrate both sides of WI
(\ref{eq:wicc}) over all space-time. Upon integration the Schwinger
$b_1$-term on the
l.h.s. vanishes. For the remaining terms on the l.h.s. two possibilities arise:
either work directly in the chiral limit or work with a non-zero quark mass
and subsequently extrapolate to the chiral limit. 
If we work directly in the chiral limit, the $(m_0 - \overline m)$ term
vanishes. The integrated $\nabla^\mu_x \hat A^f_\mu$ would also vanish, being
the integral of a four-divergence, if chiral symmetry were not broken
(absence of Goldstone bosons). In this case the WI implies a vanishing chiral
condensate. If the symmetry is broken, the presence of a Goldstone boson
guarantees a non-vanishing surface term upon integrating $\nabla^\mu_x
\hat A^f_\mu$. Thus, also the r.h.s. of the WI (i.e. the chiral condensate)
is non-zero. In the present work, in accordance with standard numerical 
computations, we will be working with a finite quark mass subsequently
extrapolated to the chiral limit. Then the surface term of the divergence
of the axial current vanishes, and the integrated WI becomes (for $f = g$)
\be
\label{eq:wiccint}
\dfrac{1}{N_f} \langle \bar \psi \psi \rangle_{sub} =
\lim_{m_0 \rightarrow m_C}
2 (m_0 - \overline m) \int d^4 x \langle P^f(x) P^f(0) \rangle 
\; .
\ee
\par
From the above expression, we can also derive two other useful identities
for the chiral condensate. They are equivalent, in theory, to the one above,
in the chiral limit and up to discretization errors. Since in practice they 
are implemented at finite quark mass and lattice spacing, they may reveal
the importance of these systematic effects. Both expressions
are obtained by writing the correlation function
$\langle P^f(x) P^f(0) \rangle$ as a time-ordered product and by inserting a
complete set of states in standard fashion. The spatial integration
$\int d^3 x$ projects zero-momentum states. Contributions from higher mass
states vanish in the chiral limit. Upon performing the time integration
$\int dt$, we find
\be
\label{eq:wip}
\dfrac{1}{N_f} \langle \bar \psi \psi \rangle_{sub} =
- \lim_{m_0 \rightarrow m_C}
\dfrac{(m_0 - \overline m)}{m_P^2}
\Big \vert \langle 0 \vert P(0) \vert P \rangle \Big \vert^2\; ,
\ee
where $m_P$ is the mass of the pseudoscalar state $\vert P \rangle$.
Note that the pseudoscalar operator without a flavour index is defined above
as, say, $P = \bar u \gamma_5 d$ (and the pion state has a $\bar d u$
flavour content). The same is true for the axial current (without flavour
index) $A_\mu = \bar u \gamma_\mu \gamma_5 d$.

Another expression for the chiral condensate is obtained by
substituting in the above equation the matrix element
$\langle 0 \vert P(0) \vert P \rangle$ by 
$\nabla^\mu_0 \langle 0 \vert \hat A_\mu(0) \vert P \rangle$
(see eq.(\ref{eq:mewi})).
The matrix element of the axial current is parameterized in terms of the
pion decay constant:
\be
\langle 0 \vert \hat A_\mu(x) \vert P(\vec p) \rangle
= i f_P p_\mu \exp(-ipx) \; ,
\label{eq:dc}
\ee
($f_P$ is the decay constant of the pseudoscalar state $\vert P \rangle$)
and thus 
\be
\nabla^\mu_0
\langle 0 \vert \hat A_\mu(0) \vert P(\vec 0) \rangle
= f_P m_P^2\; .
\label{eq:ddc}
\ee
Combining eqs.~(\ref{eq:mewi}), (\ref{eq:wip}) and (\ref{eq:ddc}) we find
\be
\label{eq:wiaa}
\dfrac{1}{N_f} \langle \bar \psi \psi \rangle_{sub} = 
- \lim_{m_0 \rightarrow m_C} \dfrac{f_P^2 m_P^2}
{4 (m_0 - \overline m)}\; ,
\ee
which is the familiar Gell-Mann--Oakes--Renner (GMOR) relation \cite{gmor}.
Note that the non-vanishing of the chiral condensate in the above equation
implies the well known linear dependence of the pseudoscalar mass squared
from the quark mass.

All determinations of the chiral condensate discussed above are subject to
taking the chiral limit ($m \rightarrow m_C$) and the continuum limit
($a \rightarrow 0$). In practice we work at non-zero quark mass $m_0$ (fixed
hopping parameter $\kappa$) and lattice spacing $a$ (fixed inverse square
coupling
$\beta$). At fixed $\beta$ the chiral limit is normally taken by extrapolating
the lattice data
to $\kappa_C$ (see comments in the introduction on the reliability of this
extrapolation and the presence of chiral logarithms). Assuming that this
extrapolation is reliable, we obtain a result which is subject to systematic
errors primarily
due to the finiteness of the lattice spacing. These effects are
$\ct{O}(a)$ for Wilson fermions. To reduce these errors, improved
actions in the spirit of Symanzik's programme \cite{sym} must be used. We have
used the tree-level improved Clover action and operators of
refs.~\cite{sw,heatlie}, which leaves us with $\ct{O}(ag_0^2)$ discretization
errors. Comparison of results obtained with the Wilson and Clover actions
enable us to estimate finite lattice spacing effects. Some technical details
on the implementation of chiral WIs with the tree-level Clover action are
given in Appendix~\ref{app:cwi}.

\section{Lattice measurements of the chiral condensate}
\label{sec:meascc}

Having shown how lattice WIs uniquely determine the subtracted chiral
condensate with Wilson fermions, in this section we will explore several
ways of measuring it from simulations. We will first discuss some
lattice details in subsect.~\ref{subs:latdet}. In subsect.~\ref{subs:naive}
we will explore several standard methods, previously used in the literature, 
for extracting the chiral condensate from lattice simulations. They are based
on eqs.~(\ref{eq:wiccint}), (\ref{eq:wip}) and (\ref{eq:wiaa}). Their main
shortcoming is that, upon passing from lattice to physical units, the
systematic error of the condensate due to the uncertainty in the lattice 
calibration is amplified. Another (minor) error amplification is due to
the extrapolation of the result from the strange quark region to the chiral
limit. In subsect.~\ref{subs:newmeas} we
will present what we consider the best method of measuring the lattice
condensate, which sidesteps these problems.
The reader uninterested in lattice technicalities may skip these subsections.
Finally, in subsect.~\ref{sub:finres} we discuss our final results.

\subsection{Computational details}
\label{subs:latdet}

In this work we have used the results of the best runs performed by the APE
group in the last few years. Both the Wilson (W) and the tree level
Clover (C) actions have been used at three values of the coupling
$\beta = 6/g_0^2$ and several quark masses in the strange quark region
(the range of the lowest-lying pseudoscalar meson in these runs is
in the range $m_K - 2 m_K$).
Table~\ref{tab:clo} shows the parameters of the runs from which we have
extracted the necessary masses and matrix elements.
A discussion of the technical details can be found in \cite{mio1} and
\cite{m_q_APE}. Here we just state that
all statistical errors have been estimated with the jacknife method
by blocking the data in groups of 10 configurations. We have checked that our
error estimates are roughly independent of the blocking size.
\setlength{\tabcolsep}{.16pc}
\begin{table}
\begin{center} 
\begin{tabular}{||c||c|c|c|c||c|c|c||}
\hline\hline       
&C60a&C60b&C62&C64&W60&W62&W64\\
\hline
$\beta$&$6.0$&$6.0$&$6.2$&$6.4$&$6.0$&$6.2$&$6.4$\\
\# Confs&490&600&250&400&320&250&400\\
Volume&$18^3\times 64$&$24^3\times 40$ &$24^3\times 64$&$24^3\times 64$&
$18^3\times 64$&$24^3\times 64$ &$24^3\times 64$
\\
\hline
$\kappa$&0.1425&0.1425&0.14144&0.1400&0.1530&0.1510&0.1488\\
        &0.1432&0.1432&0.14184&0.1403&0.1540&0.1515&0.1492\\
        &0.1440&0.1440&0.14224&0.1406&0.1550&0.1520&0.1496\\
        &  -   &   -  &0.14264&0.1409&   -  &0.1526&0.1500\\
\hline
$t_1 - t_2$ & 15-28 & 15-19 & 18-28 & 24-30 & 15-28 & 18-28 & 24-30\\
\hline
\hline
$a^{-1}(K^\ast)$&2.12(6)&2.16(4)& 2.7(1)&4.0(2)&2.26(5)&3.00(9)&4.1(2)\\    
\hline
\hline
\end{tabular}
\end{center}
\vspace{.3truecm}
\caption{Summary of the parameters of the runs with the Clover (C)
and Wilson (W) fermion actions used in this work. We also show the time
intervals in which correlation functions were fitted in order to extract
meson masses and decay constants. The bottom line is our preferred value
for the lattice spacing, obtained by fixing the mass of the vector
$K^\ast$-meson. The error is statistical.}
\label{tab:clo}
\end{table}
The APE collaboration has also performed dedicated runs at the same $\beta$
values with both the Wilson and the tree-level Clover actions, in order to
obtain NP estimates of the renormalization constants in the RI scheme
\cite{Z_bl_tli}.  The parameters of these runs and the results for the
renormalization constants of interest to the present work are reported in
table \ref{tab:ZPZA}.
More results and details on the error quoted can be found in
ref.~\cite{Z_bl_tli}. The main qualitative conclusion of ref.~\cite{Z_bl_tli}
(as seen in Table \ref{tab:ZPZA}) is that the NP method gives excellent
estimates of the renormalization constants $Z_A$ and $Z_S$, whereas $Z_P$
(and subsequently the ratio $Z_P/Z_S$) are subject to a bigger systematic
error.
\setlength{\tabcolsep}{.16pc}
\begin{table}[htb]
\begin{center}
\begin{tabular}{||l||c|c|c|c|c|c||}
\hline\hline
$\beta$ & 6.0  & 6.0    & 6.2 & 6.2    & 6.4 & 6.4  \\
\hline
Action   & C   & W & C  & W & C  & W \\
\# Confs & 100  & 100    & 180 & 100    & 60  & 60     \\
Volume   &$16^3\times 32$ & $16^3\times 32$ & $16^3\times 32$ & $16^3\times 32$
         &$24^3\times 32$ & $24^3\times 32$ \\
\hline
$\kappa$ &0.1425& 0.1530 & 0.14144 & 0.1510 & 0.1400 & 0.1488 \\
         &0.1432& 0.1540 & 0.14184 & 0.1515 & 0.1403 & 0.1492 \\
         &0.1440& 0.1550 & 0.14224 & 0.1520 & 0.1406 & 0.1496 \\
         &      &        & 0.14264 & 0.1526 & 0.1409 & 0.1500 \\
\hline\hline
$Z_S$ & 0.83(2) & 0.68(1) & 0.85(2) & 0.72(1) & 0.85(2) & 0.74(1) \\
$Z_P$ & 0.41(6) & 0.45(6) & 0.47(5) & 0.50(5) & 0.55(3) & 0.57(4) \\
$Z_P/Z_S$ & 0.49(6) & 0.66(8) & 0.56(5) & 0.69(7) & 0.65(2) & 0.77(5) \\
$Z_A$ & 1.05(3) & 0.81(1) & 1.02(2) & 0.81(1) & 1.01(1) & 0.82(1) \\
\hline
\hline
\end{tabular}
\end{center}
\vspace{0.3truecm}
\caption{Parameters of the runs used for the NP calculation of 
the renormalization constants and values of $Z_S$, $Z_P$, $Z_P/Z_S$
and $Z_A$ at a scale $\mu a \simeq 1$. All results are in the chiral limit.
The error includes both statistical and systematic effects.}
\label{tab:ZPZA}
\end{table}
\par
In lattice simulations we use dimensionless quantities; we thus define the
operators (in lattice units)\footnote{
The lack of an $1/N_f$ factor in  $\langle \bar\chi
\chi \rangle$ simply reflects the fact that, since we work in the quenched
approximation and with degenerate quark masses, there is no flavour distinction
in our numerical simulations}
\bea
\ct{P}(x) &=& a^3 P(x) \nonumber \\
\ct{A}_\mu(x) &=& a^3 A_\mu(x) \nonumber \\
\langle \bar \chi \chi \rangle &=& a^3\dfrac{1}{N_f} 
\langle \bar \psi \psi \rangle_{sub}\; .
\label{eq:lops}
\eea
The mass and decay constant of the pseudoscalar meson in lattice units are
denoted by $M_P = a m_P$ and $F_P = a f_P$. The extraction of quantities
of interest in lattice units at fixed quark mass (e.g. $M_P$, $F_P$, $a\rho$)
consists in fitting the large time behaviour of correlation functions of
suitable operators. Since this is a standard procedure we will not present it
here; the interested reader can find all relevant details of our analysis
in ref.~\cite{mio1}.
We recall that in terms of the Wilson hopping parameter $\kappa$, the bare
quark mass is given by\footnote{
Note that eq.~(\ref{eq:qmass}) is also valid in the Clover case;
i.e. even in the Clover formulation it suffers from $\ct{O}(a)$ corrections.
The corresponding $\ct{O}(a)$-improved relation will not be implemented as it
subject to some complications of a technical nature, which we fully explain
in Appendix~\ref{app:cwi}.}
\ba
\label{eq:qmass}
a(m_0 - m_C) & = & \dfrac{1}{2\kappa} - \dfrac{1}{2\kappa_C} + \ct{O}(a)\; .
\ea
\par
To pass from lattice units to physical ones, the experimental value
of several physical quantities is required (one for setting the scale and one
per quark flavour for setting the corresponding quark mass). In practice more 
than one physical quantity is used for the determination of the lattice
spacing, in order to estimate the systematic error of the lattice calibration.
In the present work we have set the scale $a^{-1}$ from the
vector meson mass $M_{K^\ast} =0.8931$~GeV. As suggested in ref.~\cite{mio1},
this gives the most reliable estimate, since it corresponds to the strange 
quark mass region, where our simulations have been carried out, and it is free
of any uncertainties due to renormalization. The result (obtained with the
methods of ref.~\cite{mio1} on the dataset of ref.~\cite{m_q_APE}) is shown in
Table \ref{tab:clo}. We see that the statistical error is about 2\%-5\%. The
systematic error on the lattice spacing has been estimated as the spread of
the $a^{-1}$ values obtained from 10 other physical quantities\footnote{
Since it is known that the scale is poorly fixed from the nucleon
mass, we have not taken this estimate into account.},
as listed in Table 11 of ref.~\cite{mio1}. This error is about 7\%. Thus, we
quote an overall 8\% error due to the uncertainties in the lattice calibration.
The discrepancies between the various lattice calibrations is believed to
reflect, at least in part, the effect of quenching.  
We will also be using some physical quantities in order to determine the
chiral limit, quark masses etc. In the region of the up and down quark masses
we have used the experimental values $m_\pi = 0.138$~GeV and
$f_\pi = 0.1307$~GeV. In the region of the strange quark mass we
have taken $m_K = 0.495$~GeV and $f_K = 0.1598$~GeV. In
the chiral limit ($m_P = 0$) we use the ``experimental'' value
$f_\chi = 0.1282$~GeV obtained by considering $f_K$ and $f_\pi$,
as functions of $m_K^2$ and $m_\pi^2$ respectively(PCAC) and by extrapolating
them naively to vanishing $m_P$.

\subsection{Standard lattice measurements of the dimensionless chiral
condensate}
\label{subs:naive}

We will now present several ``standard'' measurements of the subtracted chiral
condensate based on eqs.~(\ref{eq:wiccint}), (\ref{eq:wip}) and
(\ref{eq:wiaa}). In theory all three equations are equivalent in the continuum
limit. However, since  they are implemented in practice at fixed
lattice spacing (fixed $\beta$) and non-zero quark mass, they are subject to
systematic errors. Thus, by obtaining the chiral condensate in
several ways, we are able to study the systematic errors introduced by the
finiteness of the lattice spacing, the extrapolations to the chiral limit etc.

Once the hadronic correlation functions have been fitted and the lattice
masses and matrix elements have been extracted, we must extrapolate the
results to the chiral limit. The standard way consists in determining the
chiral point $\kappa_C$ by fitting the pseudoscalar mass $M_P^2$ as
a linear function of the quark masses
\be
M^2_{P} = C^{HS} \left( \dfrac{1}{\kappa} - \dfrac{1}{\kappa_C} \right) \; .
\label{eq:m2mq}
\ee
The slope $C^{HS}$ related to the hadron spectrum ($HS$) is also determined
from this fit.
The problem is that this relation is based on eq.~(\ref{eq:qmass}) which,
for our Clover dataset is only satisfied to $\ct{O}(a)$ and therefore it
should not be implemented for this action (see Appendix~\ref{app:cwi}). In
order to avoid this problem, we fit instead $a\rho$ as a linear function of
$M_P^2$
\be
2 a\rho  =  \dfrac{1}{C^{AWI}} M^2_{P} \; ,
\label{eq:2rhomp}
\ee
and determine the slope $C^{AWI}$ related to the Axial WIs ($AWI$). We have
verified for the Wilson data that $a\rho$ vanishes at $\kappa_C$ within our
statistical errors. 

The most straightforward way of computing the subtracted condensate is that of
eq.~(\ref{eq:wiccint}), where the term $(m_0 - \overline m)$ has been computed
from $a\rho$, with the aid of eq.~(\ref{eq:2rho}):
\be
\dfrac{\langle \bar \chi \chi \rangle}{Z_A} = \lim_{M_P^2 \rightarrow 0}
a\rho \sum_x \langle \ct{P}(x) \ct{P}^\dagger(0) \rangle \; .
\ee
Unfortunately this method is unreliable: at $\beta = 6.0$ and $6.2$ we obtained
results which are clearly incompatible with those obtained from other methods;
at $\beta = 6.4$ the statistical error overwhelms the signal. The reason is
that the correlation function $\langle \ct{P}(x) \ct{P}^\dagger(0) \rangle$,
integrated over all space-time, has a contact term at $x \approx 0$ which from
dimensional counting is seen to behave as $1/a^2$. Although its
contribution to the chiral condensate vanishes in the chiral limit (it is to be
multiplied by $(m_0 - \overline m)$), it is the dominant contribution at the 
values of lattice spacing and quark mass of the simulation. Therefore the 
extrapolation to $m_C$ becomes unreliable and with big statistical errors.
Thus, this computation of the condensate must be
discarded. This effect was first observed (with much poorer statistics
and control of the systematics) in ref.~\cite{mm}. However we note, that
in ref.~\cite{daniel} a variant of the above determination, performed
with smeared operators, gave satisfactory results in the chiral limit.
Presumably, smearing reduces the influence of the contact term on the data.

Two robust ways of computing the subtracted chiral condensate, which avoid
contact terms, are based on eqs.~(\ref{eq:wip}) and (\ref{eq:wiaa}). Once more
eq.~(\ref{eq:2rho}) is used to compute the term $(m_0 - \overline m)$. We find
for $\langle \bar \chi \chi \rangle$:
\bea
\label{eq:chi12}
\dfrac{\langle \bar \chi \chi \rangle_1}{Z_A} &=& - \dfrac{1}{2 C^{AWI}}
\lim_{M_P^2 \rightarrow 0}
\bigg \vert \langle 0 \vert \ct{P}(0) \vert P \rangle \bigg \vert ^2
\\
\dfrac{\langle \bar \chi \chi \rangle_2}{Z_A} &=& - \dfrac{1}{2} C^{AWI}
\lim_{M_P^2 \rightarrow 0} F_P^2 \; .
\nonumber
\eea
Other computations of the subtracted chiral condensate are also possible
by using eqs.~(\ref{eq:zszpm}),(\ref{eq:qmass}) and (\ref{eq:m2mq})
in order to express $(m_0 - \overline m)$ in terms of
$(m_0 - m_C)$. We avoid this option for two reasons. First, it requires
the ratio $Z_P/Z_S$ which, compared to $Z_A$, has a larger error (see results
in Table \ref{tab:ZPZA}). Second, it involves eq.~(\ref{eq:qmass}) which
cannot be implemented with our Clover data.
\setlength{\tabcolsep}{.18pc}
\begin{table}[htb]
\begin{tabular}{||c||c|c||c|c||}
\hline
\hline
Run
& $\langle\bar{\chi}\chi \rangle_1/Z_A$    
& $\langle\bar{\chi}\chi \rangle_2/Z_A$    
& $a^{-3}\langle\bar{\chi}\chi \rangle_1$    
& $a^{-3}\langle\bar{\chi}\chi \rangle_2$\\    
\hline
C60a
& 0.0040(3) 
& 0.0040(3) 
& 0.044(4)
& 0.044(4)\\   
C60b
& 0.0039(2)
& 0.0039(2) 
& 0.045(3)
& 0.046(3)\\   
W60
& 0.0045(3)
& 0.0051(4) 
& 0.048(4)
& 0.053(6)\\   
\hline
C62
& 0.0018(2)
& 0.0015(2) 
& 0.041(7)
& 0.034(5)\\   
W62
& 0.0019(1)
& 0.0016(1) 
& 0.046(5)
& 0.038(4)\\   
\hline
C64
& 0.0007(1)
& 0.0007(1) 
& 0.047(12)
& 0.049(8)\\   
W64
& 0.0008(1)
& 0.0007(1) 
& 0.053(11)
& 0.047(7)\\   
\hline
\hline
\end{tabular}
\caption{The subtracted chiral condensate determined from
eqs.~(\ref{eq:chi12}) in lattice units (second and third columns) and in
physical, GeV$^3$, units (fourth and fifth columns). An overall minus sign
has been omitted from the data. All errors are statistical.}
\label{tab:compchis}
\end{table}
\par
In Table \ref{tab:compchis} we present our results. Our comments are the
following:
\begin{itemize}
\item{The results from runs C60a and C60b, which only differ by a volume 
factor of about 1.5, are perfectly compatible. We see no finite volume
effects.}
\item{All $\langle\bar{\chi}\chi \rangle_1/Z_A$ and
$\langle\bar{\chi}\chi \rangle_2/Z_A$ results, obtained with the Wilson
action at the same $\beta$, are compatible within two standard deviations.
Those from the Clover action are compatible within one standard deviation.
This suggests that our chiral extrapolations are robust, especially in the 
Clover case where the discretization error is reduced.}
\item{The $\beta = 6.4$ results have a larger relative error for both
actions. This is probably due to the small physical volume at this value of
$\beta$. We recall that the quark mass measured at this coupling on the same
dataset in ref.~\cite{m_q_APE} also suffered from finite lattice effects. Thus
the $\beta = 6.4$ results will not be taken into account.}
\end{itemize}

Earlier works also computed the subtracted condensate at $\beta = 6.0$
with the Wilson action. The methods used were all based on eq.(\ref{eq:wiaa})
(GMOR), so they are comparable to our $\langle\bar{\chi}\chi \rangle_2/Z_A$
value:
\bea
\langle\bar{\chi}\chi \rangle_2/Z_A &=& -0.0051(4) \qquad {\rm This \,\, work}
\nonumber \\
\langle\bar{\chi}\chi \rangle/Z_A   &=& -0.006(1)\qquad {\rm Ref.~\cite{mm}}
\label{eq:cclats}
\\
\langle\bar{\chi}\chi \rangle/Z_A   
&=& -0.006(3)\qquad {\rm Ref.~\cite{daniel}} \; .
\nonumber
\eea
We see that our result is compatible with previous determinations, but
the statistical accuracy is greatly improved.

The standard method for computing the chiral condensate would now involve
renormalizing $\langle \bar \chi \chi \rangle_1$ or 
$\langle \bar \chi \chi \rangle_2$ by multiplying it with $Z_P$ (c.f.
eqs.(\ref{eq:ccs}) and (\ref{eq:cc})) and expressing it in physical units
by multiplying it with
$a^{-3}$ (c.f. eq.(\ref{eq:lops})). The latter step is the source of error
amplification, due to the cubic power of the lattice spacing. Our 8\% overall
estimate of the error due to the lattice calibration would be amplified to
24\%. This increased error is a serious shortcoming of all standard
methods of computing the condensate on the lattice.

\subsection{Lattice measurements of the physical chiral condensate}
\label{subs:newmeas}

We will now present a better method of computing the chiral
condensate on the lattice. Rather than the standard two-step approach
(i.e. first measure the dimensionless condensate in terms of dimensionless
quantities and then multiply it by $a^{-3}$) we will obtain it in physical
units, by using the physical value of the pseudoscalar decay constant
multiplied by $a^{-1}$. Thus we write the GMOR relation (\ref{eq:wiaa}) for the
renormalized condensate as follows
\be
\dfrac{1}{N_f} \langle \bar \psi \psi \rangle_1
= - \dfrac{1}{2} a^{-1} f_\chi^2 Z_S C^{HS}\; .
\label{eq:psi1}
\ee
Use of eqs.(\ref{eq:zszpm}) and (\ref{eq:m2mq}) has been made. Note that
$f_\chi=0.1282$~GeV is the ``experimental'' value in physical units (see
subsect.~\ref{subs:latdet}). The computation of the condensate based on the
above formula has several advantages. The most important is that, by
expressing it in terms of $f_\chi$, we are left with only one power of
the UV cutoff $a^{-1}$. Thus, we avoid the error amplification of the
standard methods. Another advantage is that there is no extrapolation
to the chiral limit. Instead, we only need to determine the slope $C^{HS}$
of the squared mass of the pseudoscalar meson with respect to
the quark mass. This is reminiscent of similar
procedures used in spectroscopic studies in  refs.~\cite{mio1,lm}.
Note that since we work at $\kappa$ values typical of the strange quark mass,
we are implicitly assuming that the slope will not change in the chiral region.
There is no way round this assumption, besides simulating at lighter quark
masses. Finally, another advantage of eq.~(\ref{eq:psi1}) is that the
renormalization constant is now $Z_S$, which is much better determined than
$Z_P$ with the NP method.

The only drawback of the above determination is that it cannot be used
on our Clover data as explained in Appendix~\ref{app:cwi}. The alternative
is to write eq.~(\ref{eq:wiaa}), in terms of eq.(\ref{eq:2rhomp}), as follows
\be
\dfrac{1}{N_f} \langle \bar \psi \psi \rangle_2
= - \dfrac{1}{2} a^{-1} f_\chi^2 \dfrac{Z_P}{Z_A} C^{AWI}\; .
\label{eq:psi2}
\ee
Also this determination has the advantages of having only a single power
of the inverse lattice spacing and no explicit chiral extrapolation;
a disadvantage is the larger systematic error of $Z_P$. 

Since we use the NP estimates of the renormalization constants in both
determinations of eqs.~(\ref{eq:psi1}) and (\ref{eq:psi2}), our results are
obtained in the RI scheme at a renormalization scale $a^{-1}$. Subsequently,
perturbation theory is used to NNLO to express them in the $\MSbar$
scheme (c.f. eq.~(\ref{eq:chcondm})) and run them to the conventional reference
scale of 2~GeV (c.f. eq.~(\ref{eq:rgrun})). In so doing, two further
uncertainties are introduced. The first is the $\ct{O}(\alpha_s^3)$ error
due to the truncation of the NNLO perturbative series which we consider
negligible. The second uncertainty arises because the initial
scale $\mu \simeq a^{-1}$ is known with an 8\% precision. Since the dependence
of the perturbative coefficients on the scale is logarithmic, we find that
the error introduced is at most 2\% and will be ignored.

In conclusion there are three important
errors in our results. Firstly, we have a purely statistical error of
the lattice quantities $C^{HS}$ and $C^{AWI}$. Secondly, there is the 8\%
error of statistical and systematic origin from the lattice calibration
(the $a^{-1}$ factor in eqs.(\ref{eq:psi1}) and (\ref{eq:psi2})).
Finally, we have the error in the determination of the renormalization
constants (c.f. Table \ref{tab:ZPZA}) which is also of both statistical and
systematic origin.

In table \ref{tab:ccRI} we present our results for the chiral condensate
in the $\MSbar$ scheme at 2~GeV. Our comments are as follows:
\begin{itemize}
\item{At fixed $\beta$, the two estimates from 
$\langle \bar \psi \psi \rangle_1^{\msbar}$ and
$\langle \bar \psi \psi \rangle_2^{\msbar}$
obtained with the Wilson action are perfectly compatible within the
statistical errors.}
\item{At $\beta = 6.0$, the
$\langle \bar \psi \psi \rangle_2^{\msbar}$
results obtained with the Clover
and Wilson actions are perfectly compatible within two standard deviations
of the statistical error; at $\beta = 6.2$ they are compatible within one
standard deviation of the statistical error. This implies that discretization
effects are under control.}
\item{Even within our increased statistical accuracy it is not possible to see
a systematic dependence of the chiral condensate in the $\beta$ range
6.0-6.2. The results at $\beta = 6.4$ apparently show such a dependence, but,
as we have argued in subsect.~\ref{subs:naive}, this is more likely to be a
finite size effect. Thus, we will again not take into consideration results
at $\beta = 6.4$.}
\item{As previously stated, the systematic error due to the NP renormalization
constants is small in all $\langle \bar \psi \psi \rangle_1^{\msbar}$ results
and dominant in the $\langle \bar \psi \psi \rangle_2^{\msbar}$ results.
However we cannot exclude that the very small error in $Z_S$ is underestimated.
A conservative attitude consists is regarding the larger error of $Z_P$
in the $\langle \bar \psi \psi \rangle_2^{\msbar}$ results as more realistic.}
\end{itemize}

In order to appreciate the effects of the NP renormalization on the chiral
condensate, we now present the results at $\beta = 6.2$ obtained from
the perturbative values of $Z_S$ and $Z_P$. We use the results of 
Boosted PT (BPT) (see refs.~\cite{clv,Z_bl_tli}). For the Clover action
we obtain
$\langle\bar{\psi}\psi \rangle^{\overline{MS}}_2 = - 0.0185(10)$,
whereas from the Wilson action we have
$\langle\bar{\psi}\psi \rangle^{\overline{MS}}_1 = - 0.0158(5)$ and
$\langle\bar{\psi}\psi \rangle^{\overline{MS}}_2 = - 0.0205(7)$.
It is obvious that the excellent compatibility of the corresponding results,
obtained with NP renormalization constants, has been destroyed by PT.

\setlength{\tabcolsep}{.18pc}
\begin{table}[htb]
\begin{tabular}{||c||c|c||c|c||}
\hline
\hline
Run
& $C^{HS}$
& $\langle\bar{\psi}\psi \rangle^{\overline{MS}}_1$
& $C^{AWI}$ 
& $\langle\bar{\psi}\psi \rangle^{\overline{MS}}_2$\\ 
\hline
C60a
& $ 2.98 (8) $
&   -   
& $ 3.9 (1) $
& $ 0.0141 (5)(21)$\\ 
C60b
& $ 3.04 (7) $
&  -     
& $ 4.1 (1) $
& $ 0.0146 (3)(21)$\\ 
W60
& $ 2.40 (5) $
& $ 0.0150 (3)(2)$ 
& $ 3.01 (7) $
& $ 0.0152 (4)(20)$\\ 
\hline
C62
& $ 2.9 (1) $
&  -       
& $ 3.7 (2) $
& $ 0.0147 (8)(16)$\\ 
W62
& $ 2.52(8) $
& $ 0.0156 (5)(2)$ 
& $ 2.98 (9) $
& $ 0.0158 (5)(16)$\\ 
\hline
C64
& $ 3.5(2) $
&  -       
& $ 4.2(2)  $
& $ 0.0187 (9)(10)$\\
W64
& $ 2.9 (1) $
& $ 0.0172 (7)(2)$ 
& $ 3.2 (1) $
& $ 0.0179 (8)(13)$\\ 
\hline
\hline
\end{tabular}
\caption{The pure lattice quantities $C^{HS}$, $C^{AWI}$ and the
corresponding chiral condensates (in GeV$^3$) at a scale $\mu=2$~GeV,
in the $\MSbar$ scheme.
The first error is statistical, the second comes from the NP renormalization
constants. An overall minus sign has been suppressed
in the results of the chiral condensate.}
\label{tab:ccRI}
\end{table}

\subsection{Final result}
\label{sub:finres}

From the discussion in the previous section on the results of Table
\ref{tab:ccRI} we can now derive our best estimate for the chiral condensate.
Our best results are those at $\beta = 6.2$ (high enough UV cutoff without
finite volume effects). Although all results at this $\beta$ value are
compatible, the Clover estimate is in principle preferable, as it has 
$\ct{O}(g_0^2 a)$ discretization errors. 
Thus we give as our best estimate the C62 result of Table \ref{tab:ccRI}:
\bea
\dfrac{1}{N_f} \langle \bar \psi \psi \rangle^{\msbar} \left( \mu = 2 {\rm GeV}
\right) & = &
- \left( 0.0147 \pm 0.0008 \pm 0.0016 \pm 0.0012 \right)
\,\, {\rm GeV^3} \nonumber \\
& = & - \left[ \left( 245 \pm 4 \pm 9 \pm 7 \,\, {\rm MeV} \right) 
\right]^3 \; ,
\label{eq:best}
\eea
where the first error is statistical, the second is due to the NP
renormalization and the third due to the lattice calibration. For the
first time we are able to estimate the systematic uncertainties in a
reliable way. We have also seen that our
analysis is in accordance (with improved statistical accuracy) with earlier
lattice results \cite{mm,daniel}; c.f. eq.~(\ref{eq:cclats}). The results
cited here disagree by a factor of 2 with
the recent lattice computation of ref.~\cite{rajtan}. Similar discrepancies
were seen between the quark mass estimates of this work and, say,
ref.~\cite{m_q_APE}. The two effects are not independent, since the GMOR
relation connects the chiral condensate to the quark mass.
Presumably, the results of ref.~\cite{rajtan}
are flawed by unreliable extrapolations of the lattice data obtained
at several UV cutoff values to the continuum limit.

Our result is also in agreement with the one obtained from
sum rules  (see \cite{dn} and references therein):
\bea
\dfrac{1}{N_f} \langle \bar \psi \psi \rangle^{\msbar} \left( \mu = 2 {\rm GeV}
\right) & = &
- \left( 0.014 \pm 0.002 \right) \,\, {\rm GeV^3} \nonumber \\
& = & - \left[ \left( 242 \pm 9 \right) \,\, {\rm MeV} \right]^3\; . 
\label{eq:srule}
\eea

For completeness we also give the quenched RGI result defined in 
eq.~(\ref{eq:rginv}):
\bea
\dfrac{1}{N_f} \langle \bar \psi \psi \rangle^{\rm RGI} & = & 
- \left( 0.0088 \pm 0.0005 \pm 0.001 \pm 0.0007 \right)
\,\, {\rm GeV^3} \nonumber \\
& = & - \left[ \left( 206 \pm 4 \pm 8 \pm 5 \,\, {\rm MeV} \right) 
\right]^3\; .
\label{eq:bestRGI}
\eea
The errors correspond to those of eq.~(\ref{eq:best}).

The reference scale $\mu = 2$GeV is the standard lattice choice. In order
to facilitate comparison with other determinations of the condensate,
conventionally given at the scale $\mu = 1$GeV, we run our results to the
latter scale, using NNLO perturbation theory. We find
\bea
\dfrac{1}{N_f} \langle \bar \psi \psi \rangle^{\msbar} \left( \mu = 1 {\rm GeV}
\right) & = &
- \left( 0.0124 \pm 0.0007 \pm 0.0014 \pm 0.0010 \right)
\,\, {\rm GeV^3} \nonumber \\
& = & - \left[ \left( 231 \pm 4 \pm 8 \pm 6 \,\, {\rm MeV} \right) 
\right]^3 \; ,
\label{eq:best1gev}
\eea
to be compared to the sum rule result of ref.~\cite{dn}:
\bea
\dfrac{1}{N_f} \langle \bar \psi \psi \rangle^{\msbar} \left( \mu = 1 {\rm GeV}
\right) & = &
- \left( 0.012 \pm 0.002 \right) \,\, {\rm GeV^3} \nonumber \\
& = & - \left[ \left( 229 \pm 9 \right) \,\, {\rm MeV} \right]^3\; . 
\label{eq:srule1gev}
\eea

\section{Conclusions}
\label{sec:concl}

We have computed the QCD chiral condensate from first principles in the
framework of the lattice regularization with Wilson fermions. Our result
has been obtained in the quenched approximation. We have been particularly
careful in understanding and controlling all important
sources of error (except for
quenching): (i) large field configuration ensembles have been generated in
order to minimize the statistical error; (ii) two actions (Wilson and
tree-level Clover) and several gauge couplings have been used to
control the discretization error; (iii) two lattice volumes (at one gauge
coupling) ensure some control of the finite size effects; (iv) several methods
of extracting the chiral condensate, equivalent only in the chiral limit,
give us some confidence about the extrapolations to the chiral limit;
(v) all necessary renormalizations have been performed non-perturbatively
to avoid large tadpole contributions which distort lattice
perturbative renormalization at 1-loop; (vi) the RI - $\MSbar$ renormalization
scheme matching and the RG running from the lattice scale $a^{-1}$ to the
usual scale 2~GeV have been done in continuum perturbation theory at NNLO.
Our final estimate for the chiral condensate is given in eq.(\ref{eq:best}).
It is compatible with previous lattice determinations with a much smaller
statistical error and a careful estimate of the systematic error.
It is also compatible with the sum rule result. Thus we conclude that
the chiral condensate is an order parameter of the spontaneous breaking of
chiral symmetry in QCD with massless quarks.

\begin{ack}
We are extremely grateful to J.~Gasser, V.~Gimenez, V.~Lubicz, 
G.~Martinelli, G.C.~Rossi and M.~Testa for many useful discussions.
M.T. acknowledges the support of PPARC through grant GR/L22744.
\end{ack}

\section*{APPENDICES}
\appendix
\section{The RI renormalization scheme}
\label{app:npm}

We shortly review the NP method for the renormalization constants of quark
bilinear operators. The full discussion of the method can be found in
ref.~\cite{NPM}. The most up to date results, which we use in the present work,
can be found in ref.~\cite{Z_bl_tli}.

Given a quark bilinear $O^f_{\Gamma}=\bar \psi \Gamma (\lambda^f/2) \psi$
($\Gamma$ is any Dirac
matrix), we consider the operator insertion in the quark propagator
\be
G_O(p) = \int dx_1 dx_2 \exp[i p (x_1 - x_2) ]
\langle \psi(x_1) O^f_\Gamma (0) \bar \psi(x_2) \rangle  \; ,
\ee
from which we obtain the amputated Green function computed between off-shell
quark states of  momentum $p$ in the Landau gauge:
\begin{equation}
\Lambda_O(pa)= \ct{S}(pa)^{-1}G_O(pa) \ct{S}(pa)^{-1} \; .
\end{equation}  
The above quantity is computed non-perturbatively via Monte Carlo simulations
in the Landau gauge. \cite{NPM}. Any effects from Gribov copies are small
\cite{gri}; spurious solutions \cite{leoetal} have not been
considered. The renormalization $Z_O(\mu a, g_0)$ of $O_\Gamma$
is determined by the RI condition
\begin{equation}
Z_O (\mu a)Z_q^{-1}(\mu a) \Tr\Pj_O\Lambda_O(pa) \Bigg \vert _{p^2=\mu^2}=1\; ,
\label{eq:RI}
\end{equation}
where $\Pj_O$ is a projector chosen so that the above condition is satisfied
at tree level \cite{NPM} and $Z_q$ is the wave function renormalization.
Since the projected amputated Green function can be computed non-perturbatively
with Monte Carlo simulations, eq.(\ref{eq:RI}) can be solved for $Z_O$ at a
fixed scale $\mu$ and cutoff $a$ provided that $Z_q$ is known.

In order for the RI scheme to be compatible (at large scales $\mu$) with the
WIs, $Z_q$ must be defined as \cite{NPM}
\begin{equation}
Z_q(\mu a)=\left.-i\dfrac{1}{12}Tr (
\dfrac{\partial \ct{S}(pa)^{-1}}{\partial\pslash})\right|_{p^2=\mu^2}\; .
\end{equation}
The above definition is not convenient, as it requires differentiation
with respect to the discrete variable $p$.
Thus, following ref.~\cite{NPM}, we have opted for the definition
\begin{equation}
Z'_q(\mu a)=\left.-i\dfrac{1}{12}\dfrac{Tr\sum_{\mu=1,4}\gamma_\mu \sin(p_\mu
a)\ct{S}(pa)^{-1}}{4\sum_{\mu=1,4} \sin^2(p_\mu a)}\right|_{p^2=\mu^2}\; ,
\label{eq:Z_q_WI}
\end{equation}
which, in the Landau Gauge, differs from $Z_q$ by a finite term of order
$\alpha_s^2$. The matching coefficient can be computed using continuum
perturbation theory. From refs.~\cite{Z_bl_tli,fravit} we quote
up to order $\alpha_s^2$, 
\begin{equation}
\label{eq:deltaq1}
\dfrac{Z_q}{Z_q^\prime}=
1-\dfrac{\as^2}{\left( 4\pi \right) ^2}\Delta _q^{(2)}+\ldots \; ,
\end{equation}
where
\be
\label{eq:deltaq2}
\Delta^{(2)}_q =  
 {{\left( N_c^2 - 1 \right) }\over {16\,N_c^2}} \,
 \left(3 + 22 N_c^2 - 4 N_c N_f \right) \nonumber
\ee
($N_f = 0$ in the quenched approximation).
Note that at a typical scale 
$\mu \sim 2$~GeV, $\Delta_q$ represents a tiny correction, smaller than
$1\%$. The renormalization constants used in the present work have been
corrected by this factor (c.f. \cite{Z_bl_tli}).

\section{RI/$\MSbar$ matching coefficient and RG evolution coefficient}
\label{app:match}

The matching coefficient $\Delta Z^{\RI/\msbar}$ which transforms the
renormalized chiral condensate from the RI to the $\MSbar$ scheme 
is that of the renormalization constant $Z_P$ (or $Z_S$). It has been
calculated in PT to $\ct{O}(\alpha_s^2)$ (see refs.~\cite{Z_bl_tli,fravit}
for details). The result is
\be
\Delta Z^{\RI/\msbar} = 1 + \dfrac{\alpha_s(\mu)}{4\pi} C^{(1)}
+ \dfrac{\alpha_s^2(\mu)}{(4\pi)^2} C^{(2)} + \ct{O}(\alpha_s^3)\; ,
\ee
where 
\ba
\label{eq:zmri}
C^{(1)} & = & 
  {{8 \left(N_c^2 - 1 \right) }\over {4\,N_c}} \\
C^{(2)} & = & 
  {{\left(N_c^2 -1 \right) }\over {96\,N_c^2}} \,
  \left( -309 + 3029\,N_c^2
\right.  \nonumber\\
& \phantom{x} & \phantom{xxxxxxxx}
\left. - 288\,\zeta_3 - 576\,N_c^2\,\zeta_3 - 356\,N_c\,N_f  \right)
\nonumber
\ea  
and $\zeta_3 = 1.20206\cdots$.

The evolution coefficient of the chiral condensate is derived from a
standard RG analysis. Its running is determined by the evolution of the
strong coupling $\alpha_s(\mu)$ (i.e. the Callan-Symanzik $\beta$ function)
and the scalar operator (i.e. its anomalous dimension $\gamma_S(\alpha_s)$.
In any scheme which respects vector symmetry (i.e. the vector WIs are valid)
the renormalization of the scalar operator is the inverse of the
renormalization of the quark mass\footnote{
More generally, the anomalous dimensions of the two-fermion
operators of eqs.~(\ref{eq:sp}) and (\ref{eq:va}) satisfy $\gamma_V =
\gamma_A = 0$ and $\gamma_S = \gamma_P = - \gamma_m$.}
(see for example eq.(\ref{eq:zsm})).
Thus, we can readily use the results of refs.~\cite{nnlo}
obtained in the $\MSbar$ scheme (with NDR regularization) at the NNLO.
The Callan -Symanzik $\beta$ function and the quark mass anomalous dimension
are given by
\bea
\dfrac{\beta(\alpha_s)}{4\pi} &=& \mu^2 \dfrac{d}{d\mu^2} \left(
\dfrac{\alpha_s}{4\pi} \right) = - \sum_{i=0}^{\infty} \beta_i
\left( \dfrac{\alpha_s}{4\pi} \right)^{i+2} \nonumber \\
      {\gamma_m(\alpha_s)}    &=& -2 Z_m^{-1} \mu^2 \dfrac{d Z_m}{d\mu^2}
= \sum_{i=0}^{\infty} \gamma_m^{(i)}
\left( \dfrac{\alpha_s}{4\pi} \right)^{i+1}\; ,
\eea
where
\ba
\beta _0 & = & \dfrac{11}{3}N_c - \dfrac{2}{3} N_f \\
\beta _1 & = & \dfrac{34}3N_c^2-\dfrac{10}3N_c N_f-\dfrac{(N_c^2-1)}{N_c}N_f 
\nonumber \\
\beta _2^{\msbar} & = & \dfrac{2857}{54}N_c^3+\dfrac{\left(N_c^2-1\right) ^2}{
4 N_c^2}N_f-\dfrac{205}{36}\left( N_c^2-1\right) N_f 
\nonumber \\
&\phantom{x}&\;-\dfrac{1415}{54}N_c^2N_f+\dfrac{11}{18}\dfrac{\left(N_c^2-
1\right) }{N_c}N_f^2+\dfrac{79}{54}N_cN_f^2 \nonumber\\ 
\gamma_m^{(0)} &=& 3\dfrac{N_c^2-1}{N_c}  \nonumber \\
\gamma_m^{(1)} &=& \dfrac{N_c^2-1}{N_c^2}\left( -\dfrac 34+%
\dfrac{203}{12}N_c^2-\dfrac 53N_cN_f\right)  \\
\gamma_m^{(2)} &=& \dfrac{N_c^2-1}{N_c^3}\left[ \dfrac{129}8-%
\dfrac{129}8N_c^2+\dfrac{11413}{108}N_c^4\right.   \nonumber \\
&\phantom{x}&\;\left. +N_f\left( \dfrac{23}2N_c-\dfrac{1177}{54}%
N_c^3-12N_c\zeta _3-12N_c^3\zeta _3\right) -\dfrac{35}{27}N_c^2N_f^2\right]\; . 
\nonumber  \label{eq:gam}
\ea
Note that the scheme dependence settles in only at the two-loop order for
$\gamma_m(\alpha_s)$ and three-loop order for $\beta(\alpha_s)$.
Finally, to NNLO the evolution coefficient of the chiral condensate is
\cite{nnlo}
\begin{eqnarray}
c_S^{\msbar} \left( \mu \right) &=&\as \left( \mu \right)^{\overline{\gamma }
^{(0)}_S}\left\{ 1 +\dfrac{\as}{4\pi } \left( \overline{\gamma }^{(1)}_S-
\overline{\beta }_1 \overline{\gamma }^{(0)}_S\right) \right.  \\
&+&\left. \dfrac 12\left( \dfrac{\as \left( \mu \right)}{4\pi }\right) ^2\left[
\left( \overline{\gamma }^{(1)}_S-\overline{\beta }_1\overline{\gamma
}^{(0)}_S\right) 
^2+ \overline{\gamma }^{(2)}_S + \overline{\beta }_1^2\overline{\gamma
}^{(0)}_S-
\overline{\beta }_1\overline{\gamma }^{(1)}_S-\overline{\beta }_2
\overline{\gamma 
}^{(0)}_S \right] \right\} \; ,\nonumber  
\label{eq:calfa}
\end{eqnarray}
with $\overline{\beta }_i=\beta _i/\beta _0$ and $\overline{\gamma }^i_S
=\gamma_S^{(i)}/\left(2\beta_0\right)=-\gamma_m^{(i)}/\left(2\beta _0\right)$.
The running coupling $\alpha_s(\mu)$ at NNLO in $\MSbar$ scheme is given by
\begin{eqnarray}
\label{alphaeff}
     \dfrac{\alpha_s^{\msbar}}{4\pi}(q^2) & = &
        \dfrac{1}{\beta_0 \ln(q^2)}
        - \dfrac{\beta_1}{\beta_0^3}
        \dfrac{\ln\  \ln(q^2)}{\ln^2(q^2)}
  \nonumber \\
     && + \dfrac{1}{\beta_0^5 \ln^3(q^2)}\left( \beta_1^2 \ln^2\ln(q^2)
   - \beta_1^2 \ln \ \ln(q^2)+ \beta_2^{\msbar}\beta_0 - \beta_1^2 \right)\; ,
\end{eqnarray}
where $q^2 = (\mu/\Lambda^{\msbar}_{QCD})^2$. Since we work in the 
quenched approximation $N_f=0$ in all formulae. The QCD scale in the quenched
approximation has been set to $\Lambda^{\msbar}_{QCD}=0.251 \pm 0.021$~GeV; see
ref.~\cite{capi}. 

\section{Lattice WIs with Clover fermions}
\label{app:cwi}
In this Appendix we discuss the modifications which are necessary for a
computation of the chiral condensate with the tree level Clover improved
action. This involves the discussion of several
technical details. At tree-level, the Clover action is defined as follows
\cite{sw}:
\be
S_c = S_f - a^4 \sum_{x,\mu,\nu} \dfrac{a}{4} i g_0 \bar \psi(x)
\sigma_{\mu \nu} F_{\mu \nu}(x) \psi(x)\; ,
\label{eq:sclov}
\ee
with $F_{\mu \nu}(x)$ the clover-leaf discretization of the field tensor. 
At this order all $\calo (a g_0^{2n} \ln^n a)$ terms, which
are effectively of $\calo (a)$ in the scaling limit ($g_0^2 \sim 1/\ln a$), 
are eliminated from correlation functions. At leading-log level the 
improvement of local operators can be expressed as a rotation of the 
fermion fields \cite{heatlie}:
\be
O_\Gamma^I (x) = \bar \psi^R(x) \Gamma \psi^R(x)\; ,
\label{eq:oi}
\ee
where the rotated fields are defined through
\bea 
&& \psi^R(x) = \left[ 1 - \dfrac{a\dsr}{2} \right] \psi(x)
\nonumber \\
&& \bar \psi^R(x) = \bar \psi(x) \left[1 + \dfrac{a\dsl}{2} \right]
\label{eq:psir}
\eea
and $\dsl$ and $\dsr$ are symmetric lattice discretizations of the covariant
derivatives (see ref.~\cite{heatlie} for their definition). The improved
operators $O^I$ differ from the original ones by terms proportional to the
cutoff. When $O^I$ is inserted in a Green function, these extra terms
combine with the $a^{-1}$ UV divergences to give finite contributions.
Consequently, Clover renormalization constants differ from the Wilson ones
by finite terms. 

We denote by $\ct{S}^I (x-y) = \langle \psi(x) \bar \psi(y) \rangle$ 
the quark propagator obtained by solving the Dirac equation of the Clover
action. The tree-level $\calo(a)$ improved quark propagator is
$\langle \psi^R(x) \bar \psi^R(y) \rangle$; in terms of $\cals^I(x-y)$
it is given by (see ref.~\cite{clv} for details):
\be
\langle \psi^R(x) \bar \psi^R(y) \rangle = \cals^{eff}(x-y) 
+ \dfrac{a}{2} \delta (x-y) 
+ \calo(a^2) \; ,
\label{eq:iprop}
\ee
where the effective rotated propagator $\cals^{eff}(x-y)$ is defined as
\be
\cals^{eff}(x-y) = \left[1-\dfrac{a}{2} \dsr(x)\right] \cals^I (x-y)
\left[ 1+\dfrac{a}{2} \dsl(y)\right]\; .
\label{eq:srotz1}
\ee
This is the Clover propagator we use in our computations. The reason
it contains an $\calo (a^2)$ term in its definition is that it is readily
computable in numerical simulations \cite{MARTIR}. Thus, it has been
extensively used in several Clover improved lattice QCD computations of
on-shell matrix elements. In computing correlation functions between external
on-shell hadron states, the $\delta$-function never contributes. To illustrate
the point, consider a two-point correlation function of two bilinear improved
operators $O_j^I (x) = \bar \psi^R (x) \Gamma_j \psi^R (x)$ (where $j=1,2$)
given by
\bea
\label{eq:2pfct}
&& \langle O_1^I (x) O_2^I (y) \rangle =
\langle O_1^I (x) O_2^I (y) \rangle _{n{\slashchar D}ct} \\&&
-\dfrac{a}{2} \delta(x-y) \Tr \left[ \Gamma_1 \Gamma_2 \cals^{eff}(x-y) \right]
-\dfrac{a}{2} \delta(x-y) 
\Tr \left[ \Gamma_2 \Gamma_1 \cals^{eff}(x-y) \right] \; ,
\nonumber
\eea
where 
\be
\langle O_1^I (x) O_2^I (y) \rangle _{n{\slashchar D}ct}
= -  \langle \Tr \left[ \cals^{eff}(x-y) \Gamma_2 \cals^{eff}(y-x) \Gamma_1 
\right] \rangle
\ee
(i.e. the subscript $n{\slashchar D}ct$ stands for ``no ${\slashchar D}$
contact terms"). Note that the $n{\slashchar D}ct$ correlation
functions are traces of the effective propagator $\cals^{eff}(x-y)$.
However, in the case of the chiral condensate, c.f. eq.~(\ref{eq:wiccint}),
the space-time  points of the two-point correlation function are allowed to
coincide. The on-shell matrix elements
argument is then invalidated and the complete contribution of
eq.~(\ref{eq:2pfct}) must be used with the $\delta$-function giving rise to
contact terms.

The above improvement scheme is by no means unique. Other choices of quark
field rotations are also possible, as discussed in
refs.~\cite{clv,ari,masavla}. The choice made here is not necessarily
the most convenient for improving lattice WIs. As previously pointed out,
it is however, the one implemented in all APE simulations with tree-level
Clover improvement. Thus, we need to examine how $\ct{O}(a)$-improved WIs can
be obtained for this choice of rotation (i.e. this choice of improved
operators).

Upon performing an axial variation of the improved pseudoscalar operator
$P^{Ig}(0)$ with the theory defined in terms of the Clover action, we obtain
\bea
\dfrac{\delta \langle P^{Ig}(0) \rangle}{\delta \alpha_A^f(x)} = 0
\Longleftrightarrow \nonumber \\
\label{eq:awic}
i \langle \dfrac{\delta P^{Ig}(0)}{\delta \alpha_A^f (x)} \rangle =
a^4 \sum_\mu \nabla_x^\mu \langle A^f_\mu(x) P^{Ig}(0) \rangle
- a^4 2 m_0 \langle P^f(x) P^{Ig}(0) \rangle \\
- a^4 \langle X^f(x) P^{Ig}(0) \rangle \; .\nonumber
\eea
This is eq.~(\ref{eq:awi}) with the difference that the variation
of the Clover term in the action modifies the expression for $X^f$ by a
term proportional to $a \bar \psi \sigma_{\mu \nu} \tilde F_{\mu \nu} \psi$
(c.f. eq.~(\ref{eq:sclov}); $\tilde F_{\mu\nu}$ is the dual field tensor).
The above WI is not $\ct{O}(a)$ improved for three reasons: (i) the correlation
functions contain unimproved operators; (ii) the asymmetric derivative is not
improved
($\nabla_x^\mu = \partial_x^\mu + \ct{O}(a)$); (iii) the variation of the
improved operator $P^{Ig}$ (the l.h.s. of the above) does not yield the vacuum
expectation value of the improved scalar density $\langle S^{I0}(0) \rangle$
(c.f. eq.~(\ref{eq:awi})).
The conclusion is that improved WIs cannot be obtained by the standard
procedure of performing symmetry transformations on improved operators
and the Clover improved action. Instead, we simply impose the validity of WIs
(such as that of eq.~(\ref{eq:awi})) with symmetric derivatives and
correlation functions of improved operators. Since all equations of
sect.~\ref{sec:wwi}, implemented at fixed inverse
coupling $\beta$, suffer from $\ct{O}(a)$ corrections, we can add by hand
the $\ct{O}(a)$ terms which render the correlation functions improved.
Consistency requires that the mass subtraction $\overline m(m_0)$ is also
modified by $\ct{O}(a)$ terms so that $(m_0 - \overline m^I) P^{If}$ is
$\ct{O}(a)$ improved. Consequently, the mass difference $(m_0 - \overline m^I)$
vanishes at a critical point $m_C^I$. In other words, we define $2\rho^I$ as:
\be
\label{eq:2rhoc}
2 \rho^I = \dfrac{2 (m_0 - \overline m^I)}{Z_{A^I}} =
\dfrac{\int d^3 x \overline{\nabla}^0_x \langle A_0^{If}(x) P^{If}(0)\rangle}
      {\int d^3 x \langle P^{If}(x) P^{If}(0) \rangle} \; ,
\ee
where $\overline{\nabla}^0_x$ is the symmetric lattice derivative. This
equation, with $Z_{A^I}$ computed, say, from improved WIs 
\cite{clv,gri,clvwi}
is a definition of the improved quantity $(m_0 - \overline m^I)$. With this
definition, this mass difference is renormalized by $Z_{P^I}^{-1}$ (the
improved version of eq.~(\ref{eq:zpm})).

We can now write down the improved WI for the chiral condensate. It is
the WI (\ref{eq:wiccint}) expressed in terms of improved operators and
masses. Also, contact terms must be properly taken into account according
to eq.~(\ref{eq:2pfct}). We obtain
\bea
\label{eq:wiccintc}
\dfrac{1}{N_f} \langle \bar \psi \psi \rangle ^I_{sub} &=&
\lim_{m_0 \rightarrow m_C^I}
2 (m_0 - \overline m^I) \int d^4 x \langle P^{If}(x)
P^{If}(0) \rangle \\
&=& \lim_{m_0 \rightarrow m_C^I}
2(m_0 - \overline m^I) \left[ \int d^4 x \langle P^{If}(x)
P^{If}(0) \rangle_{n{\slashchar D}ct}
-a \Tr \langle \cals^{eff}(0) \rangle \right] \; .
\nonumber
\eea
Apart from these modifications, all other expressions for the
chiral condensate (see sect.~\ref{sec:meascc}) remain valid, provided all
quantities are improved.

Similar arguments can be used in the case of vector WIs. It is not possible
to obtain Clover-improved WIs by the standard vector variations of
improved operators. Instead, we impose the validity of WI~(\ref{eq:vwi})
with all correlation functions expressed in terms of improved operators. This
implies an improved definition of the subtracted mass $(m_0 - m_C^I)$, so that
the renormalization constants $Z_{S^I}$ and $Z_m^I$ obey eq.~(\ref{eq:zsm}).
Moreover, as stressed repeatedly in this work, the simple
relationship between the quark mass and the hopping parameter of
eq.~(\ref{eq:qmass}), satisfied to $\ct{O}(a)$, needs to be improved. As this
is not a straightforward task with the rotated fields of eq.(\ref{eq:psir}),
we have avoided use of eq.(\ref{eq:qmass}) when deriving our Clover results.
This point has been explained also in ref.~\cite{m_q_APE}, where an improved
version of eq.~(\ref{eq:qmass}) valid for a different rotation of
the quark fields is given.

A final word of caution is in place here. The relationships (\ref{eq:zsm}) and
(\ref{eq:zpm}) constrain the renormalization constants involved, so that
the WIs are recovered in the continuum. These constraints relate not only
the anomalous dimensions, but also the finite parts of these renormalization
constants. The determination of the finite part of a renormalization constant
depends on the choice of lattice operator\footnote{
We have already pointed out that the definition of lattice 
operators is not unique, as higher dimensional operators, multiplied by
suitable powers of the lattice spacing, can be added at will.}
and renormalization condition. There are choices for
which the WIs are not satisfied, and consequently eqs.~(\ref{eq:zsm}) and
(\ref{eq:zpm}) are violated by finite terms. For example, in 1-loop
perturbation theory with the Clover action, $Z_m^I$ has been obtained by
renormalizing the Clover quark propagator $\langle \psi \bar \psi \rangle$ (in
the $\msb$ scheme) \cite{gabri}. Compared to the result for $Z_{S^I}$
obtained in ref.~\cite{ari}, we see that eq.~(\ref{eq:zsm}) is violated
by the finite term arising from the quark field rotations in $S^I$
\cite{m_q_APE}.
Agreement would have been reached, had the rotated propagator
$\langle \psi^R \bar \psi^R \rangle$ been used to obtain $Z_m^I$.
The discrepancy is not due to the choice of scheme, but to the choice of
an unrotated quark propagator. On the other hand, the renormalization
constants used in this work, obtained in the RI scheme with the NP method
of \cite{NPM}, are perfectly compatible with the Clover WIs.

\end{document}